\documentstyle[prd,aps,floats,tabularx,epsfig]{revtex}

\newcommand{\be}{\begin{equation}}
\newcommand{\ee}{\end{equation}}

\newcommand{\bea}{\begin{eqnarray}}
\newcommand{\eea}{\end{eqnarray}}
\newcommand{\bean}{\begin{eqnarray*}}
\newcommand{\eean}{\end{eqnarray*}}

\def\spose#1{\hbox to 0pt{#1\hss}}
\def\ltapprox{\mathrel{\spose{\lower 3pt\hbox{$\mathchar"218$}}
\raise 2.0pt\hbox{$\mathchar"13C$}}}
\def\gtapprox{\mathrel{\spose{\lower 3pt\hbox{$\mathchar"218$}}
\raise 2.0pt\hbox{$\mathchar"13E$}}}

\begin{document}
\draft
\preprint{\ 
\begin{tabular}{rr}
& 
\end{tabular}
} 
\twocolumn[\hsize\textwidth\columnwidth\hsize\csname@twocolumnfalse\endcsname 
\title{Isotropy and Stability of the Brane}
\author{M. G. Santos$^{1}$,  F. Vernizzi$^{1,2}$, and P. G. Ferreira$^{1}$}
\address{ $^1$ Astrophysics \& Theoretical Physics, University of Oxford, 
Oxford OX1 3RH, UK\\
$^2$D\'epartement de Physique Th\'eorique, Universit\'e de Gen\`eve,
24 quai E.\ Ansermet, CH-1211 Gen\`eve 4, Switzerland}
\maketitle

\begin{abstract}
We reexamine  
Wald's no-hair theorem for global anisotropy in the brane world
scenarios. We derive a set of sufficient conditions 
which must be satisfied
by the brane matter and bulk metric so that 
a homogeneous and anisotropic brane asymptotically evolves to 
a de Sitter spacetime in the presence of a positive cosmological constant 
on the brane.
We discuss the violations of these sufficient conditions and we show that 
a negative nonlocal energy density  or the presence of strong 
anisotropic stress (i.e., a magnetic field) may lead the brane to collapse.
We discuss the generality of these conditions.  
\end{abstract}
\date{\today}
] 
\renewcommand{\thefootnote}{\arabic{footnote}} \setcounter{footnote}{0}
It has been proposed that we may live confined to a four dimensional
brane in a higher dimensional spacetime \cite{branepapers}; in this
scenario the fundamental high dimensional Planck mass could be of the
same magnitude as the electroweak scale, resolving one of the
hierarchy problems in the current standard model of high-energy
physics. 

This brane world scenario generally predicts deviations from
standard model processes at high energies, a direct consequence
of the possibility that gravitons propagate in the bulk\cite{pppapers}.
Nevertheless it remains viable given that there are no
constraints on the gravitational force at very high energies
(or very small distances); indeed current constraints are
only beginning to probe sub-{\it mm} scales.
One should also expect deviations from four dimensional Einstein gravity at
high densities (or correspondingly strong gravitational fields).
A first manifestation of this was seen when deriving the
evolution equations for the cosmological scale factor, $a$, in the early
universe: it was shown that $a$ is driven
by an additional term which is proportional to $\rho^2$, where
$\rho$ is the energy density of the cosmological fluid on the brane 
\cite{binetruy}.

The two standard test situations for studying gravity in the
strongly nonlinear regime are the formation and stability of
black holes and the evolution of globally anisotropic models
in the early universe. There have been attempts at constructing
black hole solutions on the brane by extending the well-known
four dimensional solutions to higher dimensions; however, they have been 
shown
to be unstable \cite{bh}. Progress is being made in numerically evolving
the modified Einstein's equations to produce a stable analog of
the four dimensional Schwarzchild solution \cite{wiseman}. Some work has also
been done on the effect of global shear on the brane, in particular
on its effect on the dynamics of the scale factor during inflation
\cite{ssm,campos} .

One of the only unambiguous predictions of the inflationary
cosmology was put forward by Wald in 1983 \cite{waldhair}: 
it was shown that
initially expanding homogeneous cosmological models isotropize in
the presence of a positive cosmological constant. This leads
to a firm prediction: if we detect any form of global
anisotropy, then the universe cannot have undergone a
de Sitter stage. As yet there is no detection of global
anisotropy, and an upper bound on global shear was
derived from the COBE four year data \cite{bfs}. In this paper we wish
to reexamine Wald's result in the brane world scenario. The conditions
we derive are more complicated than for the four dimensional case considered
by Wald, but we can still assess their generality within the
class of cosmological fluids currently being considered in cosmology.

We will attack this problem {\em locally}, i.e., in terms of dynamical
quantities on the brane; in doing this we are ignoring possible
constraints which we may eventually set {\em on the bulk} 
due to our assumption {\em on the brane}
(we discuss the
consequences of this approach at the end of this paper).
Our starting point are the modified Einstein's equations derived
by Shiromizu, Maeda, and Sasaki \cite{sms} under the assumption of
$Z_2$ symmetry in the extra dimension (for a five dimensional bulk). 
Following the notation of Maartens \cite{m}, 
these can be written as
\be
G_{\mu \nu} =- \Lambda g_{\mu \nu} +\kappa^2T_{\mu \nu} 
+ \tilde{\kappa}^4 S_{\mu \nu} - {\cal E}_{\mu \nu},
\label{EE}
\ee
where $G_{\mu \nu}$, $g_{\mu \nu}$, 
and $T_{\mu \nu}$ are the four dimensional Einstein, metric,
and energy-momentum tensors, respectively; $\Lambda$ is the effective cosmological 
constant on the brane, and $S_{\mu \nu}$ is a term quadratic in
the energy-momentum tensor, defined by
\be
4S_{\mu \nu}=\frac{1}{3} T T_{\mu \nu} -T_{\mu \rho}T_{\ \nu}^{\rho} -
\frac{1}{2} g_{\mu \nu} \left( \frac{1}{3} T^2 - T_{\rho \sigma} 
T^{\rho \sigma} \right).
\ee
The quantities $\kappa^2 / 8 \pi$ and 
$\tilde{\kappa}^2 /8 \pi$ are the effective 
Newton constant on the brane and the {\em fundamental} 
Newton constant in the bulk, respectively.
The quadratic term $S_{\mu \nu}$ represents the matter corrections
of the effective four dimensional gravity, and is significantly
important at high energy, i.e., when 
$|T_{\mu \nu}| \gtapprox \lambda =
6 \kappa^2/\tilde{\kappa}^4$, where $\lambda$ is the brane tension.
Finally, ${\cal E}_{\mu \nu}$, which 
represents the projection of the five dimensional bulk
Weyl tensor on the brane, is symmetric and traceless, 
and transmits nonlocal gravitational degrees of freedom 
from the bulk to the brane. 
We do not make any speculation about the origin of the 
cosmological constant $\Lambda$ which we assume to be positive. 

Let us now consider two components of Eq.\ (\ref{EE}), the
``initial-value'' constraint equation,
\bea
G_{\mu \nu}u^{\mu}u^{\nu}&=& \Lambda + \kappa^2 
T_{\mu \nu}u^{\mu}u^{\nu} \nonumber \\
&+& \tilde{\kappa}^4 S_{\mu \nu}u^{\mu} u^{\nu} -
{\cal E}_{\mu \nu} u^{\mu} u^{\nu},
\label{IV}
\eea
where $u_{\mu}$ is the unit normal to the spatial homogeneous hypersurfaces,
and the Raychauduri equation,
\bea
R_{\mu \nu}u^{\mu}u^{\nu}&=&-\Lambda+\kappa^2(T_{\mu \nu}-
\frac{1}{2}g_{\mu \nu}T)u^{\mu}u^{\nu}
\nonumber \\
&+&\tilde{\kappa}^4(S_{\mu \nu}-\frac{1}{2}g_{\mu \nu}S)u^{\mu} u^{\nu}
-{\cal E}_{\mu \nu} u^{\mu}u^{\nu}.
\label{RAY}
\eea
Both $G_{\mu \nu} u^{\mu} u^{\nu}$ and 
$R_{\mu \nu} u^{\mu} u^{\nu}$ can be expressed in terms of 
the three-geometry of the homogeneous hypersurfaces 
and the extrinsic 
curvature $\Theta_{\mu \nu}=\nabla_{\nu}u_{\mu}$
of these surfaces using the 
Gauss-Codazzi equations.
For convenience we decompose $\Theta_{\mu \nu}$ 
into its trace $\Theta$ and trace-free part
$\sigma_{\mu \nu}$,
\be
\Theta_{\mu \nu} = \frac{1}{3} \Theta h_{\mu \nu} +\sigma_{\mu \nu},
\ee
where
$h_{\mu \nu}=g_{\mu \nu}+u_{\mu} u_{\nu}$ 
projects orthogonal to $u_{\mu}$,
and we rewrite Eqs.\ (\ref{IV}) 
and (\ref{RAY}) as
\bea
\Theta^2 - 3 \Lambda &=& \frac{3}{2} \sigma_{\mu \nu} \sigma^{\mu \nu}
-\frac{3}{2}  \, {}^{(3)}\!R + 3\kappa^2 T_{\mu \nu} u^{\mu} u^{\nu} 
\nonumber \\
&+& 3  \tilde\kappa^4 S_{\mu \nu} u^{\mu} u^{\nu}
- 3 {\cal E}_{\mu \nu} u^{\mu} u^{\nu}
\label{Vaibrane1}
\eea
and
\bea
\dot{\Theta} - \Lambda &+& \frac{1}{3} \Theta^2 = 
-\sigma_{\mu \nu} \sigma^{\mu \nu} 
-\kappa^2 \left( T_{\mu \nu} - \frac{1}{2} g_{\mu \nu} T \right) 
u^{\mu} u^{\nu} 
\nonumber \\
&-& \tilde{\kappa}^4 \left( S_{\mu \nu} - \frac{1}{2} g_{\mu \nu} S \right) 
u^{\mu} u^{\nu} + {\cal E}_{\mu \nu} u^{\mu} u^{\nu},
\label{Vaibrane2}
\eea
where the dot denotes $u^{\mu} \nabla_{\mu}$.
Eq.\ (\ref{Vaibrane2}) is thus the evolution equation for 
the expansion rate $\Theta$ 
and $\sigma_{\mu \nu}$ is the shear of the timelike geodesic congruence 
orthogonal to the homogeneous 
hypersurfaces.
The scalar curvature ${}^{(3)}\!R$ is given in terms 
of the structure-constant
tensor of the Lie algebra of the spatial symmetry 
group (see \cite{ellis,bianchi} for Bianchi models).

We now quickly review Wald's argument which proceeds using the well known 
energy conditions on the energy-momentum tensor $T_{\mu \nu}$. These
state that
for any non-spacelike vector $t^{\mu}$ the following inequalities hold:
\be
T_{\mu \nu} t^{\mu} t^{\nu} \ge 0
\ \ \ \mbox{and} \ \ \ \left(T_{\mu \nu} - 
\frac{1}{2} g_{\mu \nu} T\right) t^{\mu} t^{\nu} \ge 0,
\label{Strong}
\ee 
known as the weak and strong energy conditions, respectively. 

On the other hand, in standard gravity, the local quadratic
term and the nonlocal term in
Eqs.\ (\ref{Vaibrane1}) and (\ref{Vaibrane2}) are absent, and it is also well
known that for all Bianchi models, except Bianchi type-IX, 
$\,^{(3)}\!R \le 0$.
Therefore, Eqs.\ (\ref{Vaibrane1}) and (\ref{Vaibrane2}) together 
with the energy conditions  (\ref{Strong}), 
yield the inequalities
\be
\dot{\Theta} \le \Lambda - \frac{1}{3} \Theta^2 \le 0.
\ee
The second of these two inequalities 
tells us that $\Theta$ never passes through zero 
and if the 
universe is expanding at some time it will do it 
forever. Indeed, we have 
that $\Theta \ge (3 \Lambda)^{1/2}$ at all time. 
Moreover, the first inequality can be integrated and yields
\be
\Theta \le \frac{(3\Lambda)^{1/2}}{\tanh ( t\sqrt{\Lambda/3})}.
\label{Upperlimit}
\ee
Thus $\Theta$ is ``squeezed'' between the lower limit $(3\Lambda)^{1/2}$
and the upper limit in Eq.\ (\ref{Upperlimit}), which exponentially approaches 
$(3\Lambda)^{1/2}$ on a time scale $(3/\Lambda)^{1/2}$.
Again using Eq.\ (\ref{Vaibrane1}) for standard gravity (without quadratic 
and nonlocal terms), as a final result 
one obtains that
\be
\sigma_{\mu \nu} \sigma^{\mu \nu} 
\le \frac{2}{3}(\Theta^2 - 3 \Lambda) \rightarrow 0,
\ee
so that the shear of the homogeneous hypersurfaces rapidly approaches zero.
One can generalize this result to Bianchi type-IX models 
provided that $\Lambda$ is sufficiently large compared to 
spatial-curvature terms \cite{waldhair}.

If one is to extend
Wald's result to the brane world scenario, one 
must consider two new (additional) constraints that, taken together, 
can be seen as 
sufficient conditions for its validity. 
These new conditions come from the presence of the 
local quadratic term and the nonlocal term which, 
in the early 
universe, when the isotropization is supposed to take place, 
may play an important role. 
Indeed, when 
\be
{\cal E}_{\mu \nu} u^{\mu} u^{\nu} \le 0
\ee
and 
\be
S_{\mu \nu} u^{\mu} u^{\nu} \ge 0 \ \ \mbox{and}  \ \  
\left( S_{\mu \nu} -\frac{1}{2} g_{\mu \nu} S \right) 
u^{\mu} u^{\nu} \ge 0,
\label{SC2}
\ee
the r.h.s.\ of Eq.\ (\ref{Vaibrane1}) is positive and 
the r.h.s.\ of Eq.\ (\ref{Vaibrane2}) is negative,
and Wald's theorem remains valid. 
\\

Let us now separately discuss these two conditions 
and the consequence of their possible violations.
The first sufficient condition imposes a constraint on the projection
of the bulk Weyl tensor.
Using the symmetry properties of ${\cal E}_{\mu \nu}$, we can 
decompose it irreducibly as
\be
{\cal E}_{\mu \nu}=-\left( \frac{\tilde \kappa}{\kappa} \right )^4
\left[ 
{\cal U} \left(u_{\mu} u_{\nu}+\frac{1}{3}h_{\mu \nu}\right)
+{\cal P}_{\mu \nu}
+2{\cal Q}_{(\mu}u_{\nu)}\right],
\ee
where
${\cal U}$, ${\cal Q}_{\mu}$, and ${\cal P}_{\mu \nu}$ 
are the effective
nonlocal energy density, flux, and anisotropic stress arising
from the free gravitational field in the bulk, respectively.
The condition therefore becomes
that 
\be
{\cal U}\ge 0.
\ee
Note that this imposes restrictions on the dynamics of the metric
{\it off} the brane. 
As pointed out a number of times, there are no
local constraints which can fix the sign of ${\cal U}$, and this 
remains a crucial problem for general issues \cite{stability,cotrans}. 
However it was
recently shown from a global five-dimensional point of view that for all
possible homogeneous and {\it isotropic} cosmological solutions on
the brane, the bulk spacetime is Schwarzchild-AdS with the mass 
parameter of the black hole proportional to ${\cal U}a^4$. 
In this setting one has some constraints 
on ${\cal U}$ ({\em e.g.} 
${\cal U}$ is positive for closed S-AdS spacetime \cite{cotrans,bulk}). 
For anisotropic models such constraints do not exist.

We therefore choose to consider the possibility that ${\cal U}$ assumes 
whatever value and 
sign and we explore all possible brane solutions as a function of 
its value.
As a first approach we neglect the 
effect of matter on the brane: as we will show, it is straightforward 
to generalize and 
include a  
cosmological fluid on the brane, but this would obscure the simple
analysis we undertake here.
We then consider coupled equations for 
$\sigma_{\mu \nu} \sigma^{\mu \nu}$, $\Theta$, and
${\cal U}$  in order to study the importance of
the bulk term.
If we restrict ourselves to Bianchi class A models we can perform
a relatively general analysis without concerning ourselves with
off diagonal terms in the Ricci tensor. For this class of models 
we thus have ${\cal Q}_{\mu}=0$, which simplifies the set of the equations
to solve.

From the nonlocal energy conservation we have
\be
{\dot {\cal U}}+\frac{4}{3}\Theta{\cal U}+
\sigma^{\mu \nu}{\cal P}_{\mu \nu}=0,
\label{U} 
\ee
and from the Gauss-Codazzi equation we have an evolution equation
for the shear,
\be
{\dot \sigma}_{\mu \nu}+\Theta\sigma_{\mu \nu} 
+ {}^{(3)}\!R_{\mu \nu} -\frac{1}{3} {}^{(3)}\!R h_{\mu \nu}=
\left(\frac{\tilde \kappa}{\kappa}
\right)^4{\cal P}_{\mu \nu} \label{sigma},
\ee
where ${}^{(3)}\!R_{\mu \nu}$ is the Ricci tensor of 
the homogeneous hypersurfaces.
We first nevertheless restrict ourselves to models with
$\, {}^{(3)}\!R_{\mu \nu}=0$, {\em e.g.}, Bianchi type-I models. 

The system of equation obtained by combining 
Eqs.\ (\ref{Vaibrane1}), (\ref{Vaibrane2}), (\ref{U}), and (\ref{sigma})
is  not closed. This is a particular example of a more general result: 
due to the lack of an evolution equation for the nonlocal anisotropic 
stress ${\cal P}_{\mu \nu}$, the evolution of physical quantity 
on the brane cannot be predicted without making further assumptions 
on the bulk configuration \cite{sms,m}. 
We therefore choose to set  
$\sigma^{\mu \nu}{\cal P}_{\mu \nu}=0$. 
As already noted in \cite{ssm}, this condition implies 
an evolution equation for ${\cal P}_{\mu \nu}$, 
i.e., $\dot{{\cal P}}_{\mu \nu} \sigma^{\mu \nu} + (\tilde{\kappa} / \kappa)^4
{\cal P}_{\mu \nu} {\cal P}^{\mu \nu}=0$, 
which can be derived from Eq.\ (\ref{sigma}) and
is consistent on the brane, allowing us to close the system. 
We do not know whether this corresponds to a plausible physical 
form of the bulk metric, but this assumption is completely consistent
from what concerns the brane point of view.
Other assumptions under which one  can still solve the system 
are nevertheless possible and we will mention some of them 
at the end of the paper.

Solving for the nonlocal energy density ${\cal U}$ and the 
expansion rate $\Theta$, we can perform a stability analysis
of the dynamics by defining a new set of dimensionless variables:
\bea
\tau=t \left(\frac{\Lambda}{3}\right)^{1/2}, \ \ &&
X=\frac{\sigma^{\mu \nu}\sigma_{\mu \nu}}{2 \Lambda}, \nonumber \\ 
Y=\frac{\Theta}{(3\Lambda)^{1/2}}, \ \ && 
U=\frac{6 {\cal U}}{\kappa^2 \lambda \Lambda}.
\eea
This yields a second order autonomous 
system of nonlinear differential 
equations (with $'=d/d\tau$),
\bea
X'&+&6XY=0, \nonumber \\
Y'&+&Y^2+U-1+2X=0, \\ \label{simp}
U'&+&4UY=0, \nonumber 
\eea
plus a constraint equation,
\be
Y^2-X-U=1,
\label{cons}
\ee
which can be solved to find the 
asymptotic evolution of the 
${\cal U}$-term, the shear,  and the expansion.
Note that, if one considers the  
constraint equation with $U=0$, $Y^2-X-1=0$,
one recovers the evolution equations of standard gravity
for the shear and expansion rate. 

This system of equations is quite simple to analyze, and reveals
the novel behavior of the asymptotic evolution of
the brane. One can identify
three critical points: $(X,Y,U)=(0,1,0)$, which is an {\em attractor} 
and is the same fixed asymptotically stable 
point we find in standard four 
dimensional cosmology ($\Theta \rightarrow (3\Lambda)^{1/2}$, 
$\sigma^{\mu \nu} \sigma_{\mu \nu} \rightarrow 0$); 
$(X,Y,U)=(0,-1,0)$, which is a {\em repeller}, an unstable point 
in the contracting phase;
and $(X,Y,U)=(2,0,-3)$, which is a saddle point, and therefore unstable.
The latter is the most interesting point of the system since it 
corresponds to $\Theta=0$ and around this point the brane ``decides''
whether to enter the de Sitter phase or to collapse. 
\begin{figure}
\centerline{\epsfig{file= 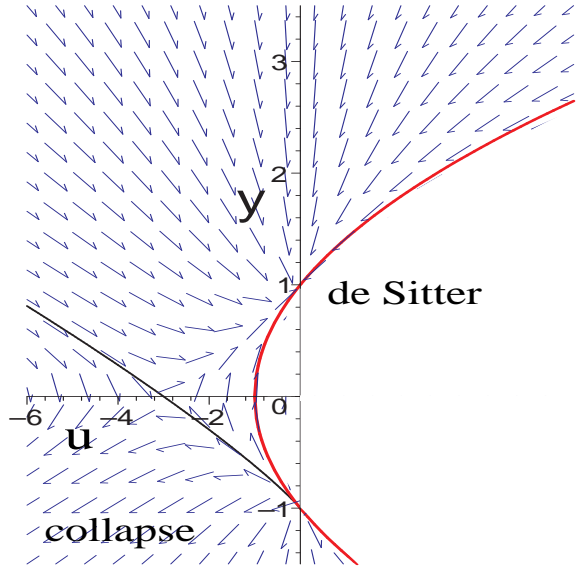, width=10cm, height=8.5cm}}
\caption{Direction field in $(U,Y)$ phase space. The thick 
(red) line represents the condition $X=0$. The thin (black) line 
separates the region of asymptotic stability from the region of instability
of the brane. For values of $U<-3$ the brane may collapse even if 
it isinitially expanding.}
\end{figure}

The orbits for this system of equations are given by 
\bea
X&=&C |U|^{3/2}, \nonumber \\ 
Y^2&=&1+U+C |U|^{3/2}, 
\label{old}
\eea 
where $C$ is an integration constant which depends on the initial conditions.
The direction field is shown in Fig.\ 1.
The thick red line 
represents the condition $X=0$. Our physical 
phase space resides where the shear is non-negative, i.e.,  
$Y^2 \ge 1+U$.
The black line passing through the saddle point represents, instead, the
{\em separatrix} between the region of asymptotic 
stability (above) and the one corresponding to instability, and is
given by
\begin{equation}
Y=\left\{
\begin{array}{cc}
-(1+U+C_0 |U|^{3/2})^{1/2},& \ \ \ U<-3 \\
\ (1+U+C_0 |U|^{3/2})^{1/2},& \ \ \ U>-3,
\end{array}
\right.
\end{equation}
where $C_0=2/3^{3/2}$.
The dynamics is therefore such that, for any initial condition 
above this line, the 
system asymptotically 
converges to $(X=0,Y=1,U=0)$, which corresponds 
to $\sigma_{\mu \nu} \sigma^{\mu \nu}=0$, 
$\Theta=\sqrt{3\Lambda}$, and ${\cal U}=0$;
however, for initial conditions below
this separatrix, 
the brane universe collapses and the anisotropy grows to infinity.

Isotropy is reached 
for any initial condition corresponding to ${\cal U} \ge 0$. 
We therefore recover the sufficient condition for the validity of 
Wald's theorem mentioned already above.
When  ${\cal U} < 0$ the brane may not isotropize and can instead collapse.
The conventional four dimensional cosmology 
resides on the subspace (the $Y$ axis)
given by ${\cal U}=0$.
As a side result, we can see that when the brane enters the de Sitter 
phase, the ${\cal U}$ term approaches zero asymptotically.
Furthermore, there is a  
region of initial conditions corresponding to small 
${\cal U}$ where it is possible to start with the 
brane contracting and still reach the stable point.
This reflects the new properties of the modified gravity and the possibility 
of avoiding the cosmological 
singularity in the presence of the nonlocal term ${\cal U}$, which 
was already noted in \cite{m}.

What we have found is not really surprising: 
the projected bulk Weyl tensor
plays a crucial role in the isotropization of the brane.
This means that there are additional constraints one must 
impose {\it in the bulk} if one is to find a complete generalisation of
Wald's theorem to the brane; in some sense this defeats
the purpose of such a no-hair theorem. 
However, 
we have been able to present a local analysis
in which we have  closed the system by simplifying the
effect of the bulk and still obtained nontrivial dynamics. 
Note that we can obtain trajectories with
${\cal U}<0$; however, this is not the crucial diagnostic for whether the
brane isotropizes or not. Of course it is of paramount
importance to see what global constraints the bulk imposes on
${\cal U}$ in the same way as has been done for homogeneous and
{\em isotropic} models.
\\

Let us go back to the second of the sufficient conditions,
Eq.\ (\ref{SC2}).
We can covariantly decompose
the brane energy momentum tensor $T_{\mu \nu}$ into its irreducible 
decomposition,
\be
T_{\mu \nu}=\rho u_{\mu} u_{\nu}+ph_{\mu \nu}+\pi_{\mu \nu}+
2q_{(\mu}u_{\nu)},
\label{tau}
\ee
where 
$\rho$ and $p$ are the energy density and isotropic pressure, 
$q_{\mu}$ is the energy flux,
and $\pi_{\mu \nu}$ is the anisotropic stress of the matter on the brane.

The restrictions 
for the quadratic term therefore becomes
\be
2\rho^2-3\pi_{\mu \nu}\pi^{\mu \nu}\ge0 \ \  \mbox{and} 
\ \ 2\rho^2+3\rho p-3q_{\mu}q^{\mu}\ge0.
\ee
These conditions are trivially satisfied  for the case of a perfect fluid 
(which satisfies
the  energy conditions) and for a free homogeneous scalar field
with a velocity flow coincident with $u^{\mu}$. 
However
they may not be satisfied in general. 

An unambiguous example of violation of one of the above conditions is given 
by a cosmological pure magnetic field $B^{\mu}$, such that $B_{\mu}u^{\mu}=0$ 
(for an extensive list of anisotropic fluids and sources see \cite{barrow});
in this case we have that
\bea 
2\rho_{B}^2-3\pi_{\mu \nu}\pi^{\mu \nu}=
-\frac{3}{2}B^4 &<& 0, \nonumber \\
2\rho_{B}^2+3\rho_{B} 
p_{B}-q_{\mu}q^{\mu}=\frac{3}{2}B^4 &>& 0, \nonumber
\eea
where 
\be
\rho_{B}=3 p_{B}=B^2/2
\ee 
and 
\be
\pi_{\mu \nu}=-B_{\mu}B_{\nu}+
(1/3) h_{\mu \nu} B^2.
\ee

For simplicity we restrict ourselves to the case of 
Bianchi type-I cosmology, which can naturally accommodate a spatially 
homogeneous ordered magnetic field,
and we neglect any other form of matter.
The presence of the magnetic field introduces a preferred direction
and  we can consider  the possibility that
the background anisotropy is due to this presence.
This corresponds to  the typical  
situation in which the magnetic field is
a shear eigenvector,
\be
\sigma_{\mu \nu} B^{\mu} = \sigma_{\bf z} B_{\nu},
\label{sheareigen}
\ee
which does not represent the most general case; however it 
considerably simplifies 
the equations \cite{tm}.
We consider the back-reaction from the bulk
but, to be able to close the system, we assume 
$\sigma^{\mu\nu}P_{\mu\nu}=0$ and $P_{\mu\nu}B^{\mu}=0$.
In this case, together with the 
Raychauduri equation, the full set of equations to solve becomes
\be
\dot{\rho}_{B} + \Theta (\rho_{B} + p_{B} )
+ \sigma^{\mu \nu} \pi_{\mu \nu} =0
\ee
from the local energy conservation,
\bea
\dot{\cal U} + \frac{4}{3} \Theta {\cal U} 
&=& \frac{\kappa^4}{24} [ 6 \pi_{\mu \nu} \dot{\pi}^{\mu \nu} 
+ 6(\rho_B +p_B) \sigma^{\mu \nu} \pi_{\mu \nu}  \nonumber \\
&+&
2 \Theta \pi^{\mu \nu} \pi_{\mu \nu} -
2 \sigma ^{\mu \nu} \pi_{\alpha \mu} \pi^{\alpha}_{\ \nu}]
\eea
from the nonlocal conservation equation, and
\bea
\sigma_{\mu \nu} {\dot \sigma}^{\mu \nu} + 
\Theta \sigma_{\mu \nu} \sigma^{\mu \nu}= \kappa^2 \pi_{\mu \nu}
\sigma^{\mu \nu} &-&
\frac{ \tilde{\kappa}^4}{12} [ (\rho_B +3p_B) \pi_{\mu \nu} \nonumber
\\ 
&-& \pi_{\alpha \mu}\pi_{\nu}^{\ \alpha}]\sigma^{\mu \nu}
\eea
from the Gauss-Codazzi equation.

Defining the additional new variables
\be
W= \sigma_{\bf z}
\left(\frac{3}{\Lambda}\right)^{1/2} \ \ 
{\mbox {\rm and}} \ \ 
Z=\frac{\rho_B}{\lambda}  \nonumber
\ee
we obtain an autonomous nonlinear system,
\bea
X'&+& 6YX+4\chi W[Z- (4/3) Z^2]=0 \nonumber, \\
Y'&+& Y^2-1+2X+2\chi [Z+(3/2)Z^2]+U=0 \nonumber, \\
Z'&+& 4ZY-2WZ=0, \label{long} \\
U'&+& 4UY-4 \chi Z^2[(5/3)W-6Y]=0 \nonumber, \\
W'&+& 3YW+8\chi [Z-(4/3)Z^2]=0, \nonumber
\eea
together with the constraint equation
\be
Y^2-2\chi[Z-(3/2)Z^2]-U-X=1.
\ee
Here $\chi= \kappa^2 \lambda / 2 \Lambda$.

This system has at least three critical points   
which correspond to those obtained 
analyzing the case without magnetic field: 
$(X=0, Y=1, Z=0, U=0, W=0)$ is
the de Sitter attractor,
$(X=0, Y=-1, Z=0, U=0, W=0)$ is the repeller, and
$(X=2, Y=0, Z=0, U=-3)$ is a saddle point for any value of 
$W$. They are all at $Z=0$ and, indeed,
$\rho_B=0$ is an {\em invariant submanifold} of the system,
i.e., if an orbit starts with $\rho_B=0$ it will
maintain this condition.
When the magnetic field is much weaker than the brane tension
$Z \ll 1$, we can neglect all the $Z^2$ contributions. In this
case we still find a region
of initial conditions corresponding to collapse of the brane,
but only for $U<-3$; this means that the 
negativity of the ${\cal U}$-term is responsible for the collapse 
and not the magnetic field.

When neglecting the $Z^2$ contribution we can 
eventually set $U=0$ and recover standard gravity.
Note that this is {\em not} possible when the quadratic term
is included. Indeed,
by analyzing the full system of differential equations one finds that
$U=0$ is not a consistent solution. 
By setting $U$ and its derivative to zero we have a conservation equation
involving quadratic terms of the energy momentum tensor and implying
$W=(18/5)Y$, which is not consistent with the evolution equations 
for  $Y$ and $W$.
We  therefore find an interesting result: in this setup
we cannot set the nonlocal energy density to zero unless we neglect 
quadratic terms. It is
conceivable that this is a consequence of discarding the effect of
the nonlocal anisotropic stress, and we are aware that 
without a full analysis of the
bulk to completely solve the evolution equations  this
must remain an open problem in this work.
This example, nevertheless, 
underline the importance of the back-reaction
of nonlocal bulk effects in the presence of
an anisotropic stress on the brane.
(As a perturbation, the effect of anisotropic stresses was 
considered and discussed in \cite{carsten}.)

When $\chi=0$ the evolutions of $X$, $Y$, and $U$ are  decoupled 
from $Z$ and $W$, and the phase space is given by the product
$(Z,W) \times (X,Y,U)_{Z=0}$.
The existence of other critical points depends on the
value of the parameter $\chi$. 
For 
\be
0 \le \chi  < \frac{16}{27},
\ee
there are three more saddle points, one at $(X=2-(27/8)\chi, Y=0, Z=3/4, 
U=-3 + (57/16)\chi,W=0)$, and other two at $Y \ne 0$.  
These saddle points degenerate to 
one when $\chi=16/27$ at the {\em bifurcation} 
point $(X=0, Y=0, Z=3/4, U=-3+(57/27),W=0)$. 
For $\chi > 16/27$ there is no other saddle point.

The novelty in the behavior of this system is the fact that
there are regions of phase space where the brane is unstable
even when the ${\cal U}$-term is positive. 
These regions can be found by numerically integrating system 
(\ref{long}) and they typically appear for large
value of the magnetic field $\rho_B \gtapprox \lambda$.
Therefore   the introduction of the anisotropic stress 
present in the quadratic term of the projected Einstein's equations 
makes the brane unstable even when the ${\cal U} \ge 0$ condition is satisfied.
We conclude that the  violation of the second sufficient 
condition for the validity of Wald's theorem 
is enough to cause 
an expanding brane to collapse, provided that the source of 
violation is sufficiently strong.

\begin{figure}
\centerline{\epsfig{file= 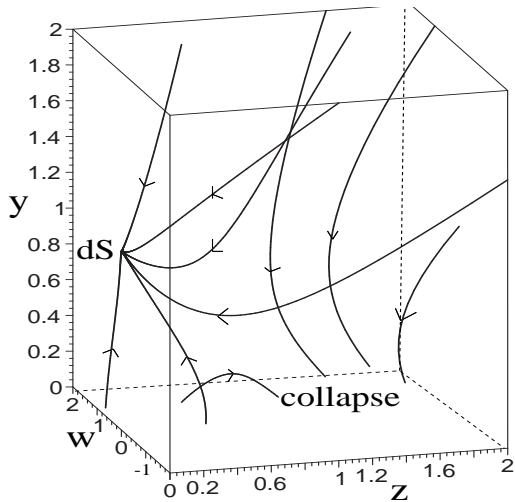, width=8cm, height=7cm}}
\caption{Phase space corresponding to the invariant submanifold $W^2 = 4X$.
For large values of $Z$, typically $Z \gtapprox 1$, the brane can collapse, 
even if $U$ is positive. Note that, since $W^2 = 4X$, 
this phase space cannot contain the new saddle point at $Z=3/4$.}
\end{figure}
In order to visualize the five-dimensional phase space 
and have an idea of the orbits
of the system 
we use the fact that $W^2 = 4X$ is another invariant submanifold.
It corresponds to the axial symmetry of the brane universe 
around the ${\bf z}$ direction and 
is realized when  
$\sigma_{\mu \nu} \propto \pi_{\mu \nu}$.         
This condition
allows us to get rid of the evolution equation for 
$X$ and use 
the constraint equation to eliminate $U$.
We are thus left with a closed system of three differential 
equations in $Z$, $W$, and $Y$.
In Fig.\ 2 we plot some typical trajectories of this system for $\chi=1/2$.
As one can see, for large value of the energy density of the 
magnetic field, $Z \gtapprox 1$,  
the brane can collapse.
\\

Let us discuss some possible generalizations of our results.
In the analysis undertaken so far, we 
have assumed that the curvature of the homogeneous spatial
hypersurfaces is zero. 
However one can still generalise some of the results 
obtained for the case $\rho_B=0$,
by including the curvature term in the
system of equations (\ref{simp}). 
Defining the dimensionless variable
\be
R= \frac{\, {}^{(3)}\!R}{\Lambda},
\ee
the relevant equations become
\bea
Y'&+&Y^2+U-1+2X=0, \\ \nonumber
U'&+&4UY=0, \nonumber
\eea
with the constraint
\be
Y^2-X-U-1=-R/2 \ge 0, \label{nnew}
\ee
where we have used  that $\, {}^{(3)}\!R \le 0$. This, as already stated, 
is true for all Bianchi
models except type-IX.

Let us define  with $Y_{R=0}(U)$ the ``old'' 
orbits of the system obtained by 
setting the curvature term to zero, and
given by Eq.\ (\ref{old}).
By carefully integrating the evolution equations for $Y$ and $U$ and using the
inequality (\ref{nnew}), 
one obtains the following behavior of the ``new'' orbits $Y(U)$:
if at some point of the phase space of Fig.\ 1
$Y(U(0)) = Y_{R=0}(U(0))$ (a new orbit crosses an old one at 
$\tau=0$),
then 
\bea
Y(U(\tau)) &\ge& Y_{R=0}(U(\tau)), \quad  \tau  > 0, \nonumber \\
Y(U(\tau)) &\le& Y_{R=0}(U(\tau)), \quad  \tau < 0. \nonumber 
\eea
By again considering Fig.\ 1 and applying 
these inequalities,
we can easily conclude 
that the region of asymptotic stability 
is enlarged by the introduction of the curvature term.
Note that the orbits never cross  the $Y$ axis since there $U'=0$.
Indeed  we still find the attractor 
and the repeller at the usual positions $(X,Y,U)=(0,\pm 1, 0)$, 
corresponding to
$R=0$.
However, it is impossible to show the existence of a saddle point
at $Y=0$ unless one can 
show that $R$ is stationary at this point. 
This is true for models in which the {\em surfaces of transitivity} 
are isotropic 
(see \cite{ellis}),
\be
{}^{(3)}\!R_{\mu \nu}=\frac{1}{3} \, {}^{(3)}\!R h_{\mu \nu}, 
\ee 
but is not generally the case.
Looking at Fig. 1, in these models the saddle point is displaced 
towards the left, at $U=-3+R$.

According to these inequalities it is still possible 
that the universe recollapses. Other types of non de Sitter behavior
are of course nevertheless possible, in particular, an 
asymptotic approach to an 
Einstein static brane. However, even with the introduction of a
curvature term, it seems reasonable to conjecture that the only 
types of stable asymptotic behaviors are an asymptotic approach to de Sitter 
spacetime and recollapse, as in the case of ${}^{(3)}\!R_{\mu \nu}=0$.
It is well known that the Kasner (vacuum Bianchi type-I model) 
collapsing solution is an attractor of Bianchi class A models with ordinary 
matter content. It would of course be important to understand if 
this is still the case in the presence of a nonlocal bulk effect.
We reserve this study for future work.

The effect of matter and other possible
cosmological fluids has been discarded in our analysis,
but it can be included quite straightforwardly 
by adding a perfect fluid moving orthogonally to the homogeneous 
hypersurfaces
with state equation $p = \omega \rho$
to our system of equations.
The presence of a new variable $\rho$ 
introduces an extra dimension in the phase space and
a new parameter $\omega$.

When the magnetic field is not taken into account
we can integrate the orbits of the system
using the dimensionless variable $Z_F = \rho/\lambda$.
This yields
\bea
X &=& C |U|^{3/2},  \nonumber \\
Z_F &=& D |U|^{3(\omega+1)/4},  \nonumber \\
Y^2&=&1+U+C |U|^{3/2}  \nonumber \\
&+&\chi (2D |U|^{3(\omega+1)/4}+ D^2|U|^{3(\omega+1)/4}), \nonumber
\eea
where $D$ is a new integration constant.
The attractor 
and the repeller are at the usual positions $(X,Y,U)=(0,\pm 1, 0)$
and $Z_F=0$. However, now the  saddle point at $Y=0$ 
becomes a line which depends on $Z_F$:
\be
U=-3( 1-\chi(\omega-1)Z_F - \chi \omega Z^2_F ).
\ee
Note, however, that for $\omega > -1/3$ 
the allowed values of $Z_F$ of this line are 
limited by the constraint $X \ge0$.
The separatrix between the region of asymptotic stability and instability
is the two dimensional surface which forms the saddle line at the intersection with
$Y=0$. The region of instability always lies below this separatrix. 
The general characteristics of the phase space of the system, as 
illustrated in Fig.\ 1, remain unchanged: the only two stable asymptotic 
behaviors are an approach to the de Sitter spacetime and recollapse; 
recollapse can only 
occur when 
${\cal U}$ is negative. 
On the $U=0$ plane, we recover the results of \cite{campos}.

In order to close the system of equations discussed above, we have 
always chosen to neglect the nonlocal anisotropic stress ${\cal P}_{\mu \nu}$.
As already discussed, this always corresponds to a consistent solution 
in the bulk, provided that the equations are consistent on the brane, although
it may not correspond to some plausible physical bulk configuration. 
For future investigations we mention three  other possibilities which 
allow one to consistently close the system on the brane.
If you have some sort of anisotropic stress on the brane 
a very simple possibility is of course to assume 
that $\dot{{\cal U}}={\cal U}=0$. In this case one 
obtainss a consistent closed
system provided that a nonzero ${\cal P}_{\mu \nu}$ is considered.

A somewhat more nontrivial assumption is the following. 
Since the nonlocal anisotropic stress ${\cal P}_{\mu \nu}$ is the spatial 
component of the projected five-dimensional bulk Weyl tensor 
on the brane, it is physically plausible to have 
a proportionality between this term and
the electric part of the brane Weyl tensor,
\be
\alpha \left( \frac{\tilde{\kappa}}{\kappa} \right)^4 
{\cal P}_{\mu \nu}=  E_{\mu \nu} = 
 C_{\mu \alpha \nu \beta} u^{\alpha} u^{\beta},
\ee
where $C_{\mu \alpha \nu \beta}$ is the Weyl tensor on the brane.
With the help of the spatial traceless component of Einstein's equations 
[see \cite{m}, Eq.\ (A3)] one
can find $E_{\mu \nu}$ in terms of the shear and the Ricci tensor 
of the homogeneous hypersurfaces and have an expression for  
${\cal P}_{\mu \nu}$ which depends on a constant parameter $\alpha$.

Another possibility to explore is to assume that 
${\cal P}_{\mu \nu}$ evolves according to a ``free'' evolution equation,
\be
\dot{{\cal P}}_{\mu \nu} + \beta \Theta {\cal P}_{\mu \nu}=0,
\ee
where $\beta$ is a constant parameter. 
In both cases ${\cal P}_{\mu \nu}$ feeds into 
the propagation equation for the shear and the ${\cal U}$-term, 
possibly giving rise to interesting dynamics, which can be studied by varying 
$\alpha$ and $\beta$.
\\

We briefly summarize the present work. We have derived a set of 
sufficient conditions to be satisfied by the local brane matter and the 
nonlocal energy density, so that a homogeneous and anisotropic 
brane with tension $\lambda$,
satisfying modified Einstein's equations with a 
positive effective 
cosmological constant,
asymptotically evolves to de Sitter spacetime. 
We have discussed the violation of these conditions, showing that,
in the presence of a negative nonlocal energy density ${\cal U}<0$ 
and/or strong anisotropic stress 
$|\pi_{\mu \nu}| \gtapprox \lambda$, a Bianchi type-I brane, 
even if initially expanding, is unstable and may collapse. 
This is our main result.
We have also given an explicit example of source of anisotropic stress 
by considering the case of a cosmological magnetic field on the brane,
and we have shown that the back-reaction of nonlocal effects cannot be 
neglected. We have discussed possible generalisations.

Finally, let us reemphasise a key aspect of our analysis. We have studied
the asymptotic behavior of the brane from a {\em local} point of view. 
On the one hand, this approach is incomplete and this is
evident in what we have found: the bulk back-reaction plays a crucial
role in the evolution of the brane. This back-reaction has been
discarded in previous local analyses where isotropic fluids were
considered [this is in fact consistent with our analysis, merely
corresponding to a choice of integration constant when solving
Eqs.\ (\ref{simp})]. Given that we do not have an evolution
equation for the nonlocal anisotropic stress 
${\cal P}_{\mu\nu}$, we have only considered a subspace
of possible solutions such that we could close the system. 
We should point
out, however that the complementary approach, i.e., constructing complete
bulk solutions and assessing their effect on the brane, may also
be restrictive
although the two approaches are clearly necessary for
a complete understanding of the asymptotic stability of the brane.
\\

{\it Acknowledgements}: We thank 
C. Van de Bruck, M. Bruni, A. Campos, M. Depken, 
R. Durrer, K. Kunze, J. Levin, R. Maartens, C. F. Sopuerta, and  
T. Wiseman for useful discussions. F.V. acknowledges support
from the Swiss NSF and MC Training Site Grant 
and P.G.F. from the Royal Society. 
M.G.S. was supported by the FCT under program PRAXIS XXI/BD/18305/98.

\tighten

\end{document}